\documentclass{amsart}
\usepackage{color}
\usepackage{graphicx}
\usepackage{amsmath,amscd}
\usepackage{amssymb}
\input epsf
\usepackage{epsfig}

\usepackage{enumerate}

\theoremstyle{definition}

\theoremstyle{remark}

\numberwithin{equation}{section}
\newcommand{\Z}{\mathbb Z}

\newcommand {\hide}[1]{}

\begin{document}
    \title[Key-agreement based on automaton groups]
    {Key-agreement based on automaton groups}
    {
        \let\thefootnote\relax\footnote{2010 Mathematics Subject Classification 15A80, 94A60}
    }
    \author{Rostislav Grigorchuk}
    \address{Mathematics Department, Texas A \& M University, College Station, TX 77843-3368, USA}
    \email{grigorch@math.tamu.edu}
    %\thanks{}
    \author{Dima Grigoriev}
    \address{CNRS, Math\'ematiques, Universit\'e de Lille, Villeneuve d'Ascq, 59655, France}
    \email{Dmitry.Grigoryev@univ-lille.fr}

    \begin{abstract}
We suggest several automaton groups as platforms for
Anshel-Anshel-Goldfeld key-agreement metascheme. They include
Grigorchuk and universal Grigorchuk groups, Hanoi 3-Towers group,
the Basilica group and a subgroup of the affine group $Aff_4(\Z)$.
 \end{abstract}

    \maketitle

\section*{Introduction}

Typically abelian groups are involved in cryptography, say in RSA and Diffie-Hellman schemes
(see e.~g. \cite{Menezes}, \cite{Myasnikov} and the references there). But they are vulnerable with respect to quantum machines. Thus, for
post-quantum cryptography one tries to use non-abelian groups (some attempts one can find in
 e.~g.  \cite{GP}, \cite{GS}, \cite{Kahrobaei} and in the references there). In this paper we suggest several groups $G$ as candidates for
 platforms for Anshel-Anshel-Goldfeld key-agreement metascheme \cite{AAG} (section~\ref{Anshel}).

To break Anshel-Anshel-Goldfeld scheme over a group $G$ an adversary needs to solve a system of simultaneous conjugacies of the form
 $xu_ix^{-1}=v_i,\, 1\le i\le m$ for given $u_i,v_i\in G,\, 1\le i\le m,\, a_1,\dots,a_n\in G$ and unknown $x\in \langle a_1,\dots,a_n\rangle$.
 On the other hand, to perform a communication between Alice and Bob via a public channel, the word problem in $G$ should have a small (say,
 polynomial) complexity. We suggest some automaton groups \cite{Grigorchuk}, \cite{Bartholdi}, \cite{Nekrashevich} (see section~\ref{Automaton}) as $G$ for which the word problem is known to have the polynomial complexity
%$O(n\log n)$.
The conjugacy problem for automaton groups was studied in
\cite{Zapata},  \cite{Wilson}. Observe that automaton groups are
convenient for algorithmic representation.

In section~\ref{Grigor} we consider Grigorchuk group \cite{Grigorchuk}. Note that in \cite{Leonov}, \cite{Rozhkov} the algorithms (without complexity analysis) for the conjugacy problem in Grigorchuk group were proposed, later in
\cite{Lysonok} a polynomial complexity algorithm for the conjugacy problem in Grigorchuk group was exhibited. But the problem of simultaneous conjugacies looks difficult in Grigorchuk group. Also mention that there was an attempt to use Grigorchuk group in cryptography in a different way \cite{Garzon} which was later broken \cite{Petrides}. In section~\ref{basilica} we discuss the Basilica group \cite{Zuk01} which is defined by an automaton with 3 states. In section~\ref{universal} we consider the universal Grigorchuk group \cite{Grigorchuk}, \cite{Nagnibeda}. In section~\ref{Hanoi} we discuss the group of Hanoi Towers on 3 pegs \cite{Sunik}. Finally, in section~\ref{unsolvable} we consider a subgroup of the affine group $Aff_4(\Z)$ with the unsolvable conjugacy problem.

\section{Anshel-Anshel-Goldfeld key-agreement metascheme}\label{Anshel}

We recall the key-agreement scheme from \cite{AAG} (cf.  \cite{GP} where its extension to multiparty communications is exhibited, also  \cite{Ushakov}
). Let $G$ be a group and
$a_1,\dots,a_n,\, b_1,\dots,b_m\in G$ be some publically given elements. Alice chooses her private element $a=a_{p_1}\cdots a_{p_s}\in \langle a_1,\dots,a_n \rangle$, while Bob chooses his private element $b=b_{q_1}\cdots b_{q_t}\in \langle b_1,\dots,b_m \rangle$. Alice transmits (via a public channel) elements $a^{-1}b_ia,\, 1\le i\le m$, while Bob transmits $ba_jb^{-1},\, 1\le j\le n$. After that Alice computes
$$bab^{-1}=ba_{p_1}b^{-1}\cdots ba_{p_s}b^{-1},$$
\noindent while Bob computes
$$a^{-1}ba=a^{-1}b_{q_1}a\cdots a^{-1}b_{q_t}a.$$
\noindent Finally, the commutator $a^{-1}(bab^{-1})=(a^{-1}ba)b^{-1}$ computed by both Alice and Bob, is treated as their common secret key.

So, an adversary has to find $A\in \langle a_1,\dots,a_n \rangle,\, B\in \langle b_1,\dots,b_m \rangle$ such that $A^{-1}b_iA=a^{-1}b_ia,\, 1\le i\le m$ and
$Ba_jB^{-1}=ba_jb^{-1},\, 1\le j\le n$ (note that the right-hand sides of the latter equalities are known). Then one can verify that $a^{-1}bab^{-1}=A^{-1}BAB^{-1}$. We emphasize that an adversary has to search a solution $A$ of the problem $A^{-1}b_iA=a^{-1}b_ia,\, 1\le i\le m$ in the subgroup $\langle a_1,\dots,a_n \rangle$ which makes the task even harder than the customary \emph{simultaneous conjugacy problem}.
Thus, our goal is to exhibit groups with the polynomial complexity of the word problem and difficult problem of solving systems of conjugacies.

We produce several candidates for such groups among automaton groups
(see e.~g. \cite{Bartholdi}, \cite{Grigorchuk},
\cite{Nekrashevich}).

\section{Automaton groups}\label{Automaton}

Denote by $X=\{0,\dots,k-1\}$ an alphabet and by $S$ a finite set that we will call a set of the states. An automaton of Mealy type on $X$ with a set $S$ of states is defined by a transition function $\tau : S\times X\to S$ and an output function $\pi:S\times X\to X$. If for each $s\in S$ the function $\pi(s,\cdot)\in Sym(k)$ is a permutation then the automaton is called invertible.

Denote by $T$ a rooted $k$-regular tree and by $T_0,\dots,T_{k-1}$, respectively, the rooted subtrees of $T$ with their roots in the children of the root of $T$. The paths (without back tracking) in $T$ starting at its root correspond to the words in the alphabet $X$. Denote by $X^l$ the set of the words of the length $l$ over $X$, by $X^*$ the set of all the words, and by $X^{\infty}$ the set of all the infinite words over $X$. Each
state $s\in S$ provides an action on $T$ being its automorphism: $s$ acts by a permutation $\pi(s,\cdot)$ on the roots of subtrees $T_0,\dots,T_{k-1}$, and in its turn $s$ acts recursively as $\tau(s,i)$ on the subtree $T_i, 0\le i <k$.

Thus, for an invertible automaton $A=(S,X,\tau,\pi)$ this defines a
group $G(A)$ of automatically defined automorphisms of $T$ with the
operation of composition. The group $G(A)$ (see e.~g,
\cite{Bartholdi}, \cite{Grigorchuk}, \cite{Nekrashevich}) is
generated by the words over $S\cup S^{-1}$ where for the state
corresponding to $s^{-1}$ the permutation
$\pi(s^{-1},\cdot)=(\pi(s,\cdot))^{-1}$ and
$\tau(s^{-1},i)=(\tau(s,i))^{-1}$. We refer to the length $|g|$ of
an element $g\in G(A)$ as its length in the generators $S\cup
S^{-1}$ (clearly, the length depends on a representation in the
generators, we'll be interested in upper bounds on the length, so no
misunderstanding would emerge). For an element $g\in G(A)$ we define
its portait (see e.~g. \cite{Bartholdi}, \cite{Grigorchuk}) of a
depth $d$ as the collection of the following data: a permutation of
the action of $g$ (denoted by $g_x$ on $X^d$ and for every word
$x=x_1\cdots x_d\in X^d$ the action of $g$ on the subtree $T_x$ of
$T$ with the root $x$ (being an element of $G(A)$, these elements
are called \emph{sections}).

In all the examples of automaton groups $G(A)$ considered below
(except for the last one) two elements $g_1,g_2\in G(A)$ are equal
iff their portraits of the depth $\log(|g_1|+|g_2|)$ coincide.
Moreover, the sections of all the words of this length over $X$ have
constant size $O(1)$ (we'll refer to it as the \emph{portrait
property}). This is due to the \emph{contracting property}
established for the groups $G(A)$ considered below (except for the
last one): there exist $\lambda <1, c,l$ such that $|g_x|<\lambda
|g|+c$ for all $g\in G(A),\, x\in X^l$. The contracting property
immediately  allows one to solve the word problem in $G(A)$ within
the polynomial complexity. On the other hand, it seems that the
problem of solving a system of simultaneous conjugacies is difficult
in all the automaton groups under consideration, the key-agreement
scheme from section~\ref{Anshel} based on any of these groups looks
hard to be broken.

Thus, one can compute the portrait within the polynomial complexity, and the portrait (or its binary encoding) will be used as a common secret key by Alice and Bob.

\subsection{Grigorchuk group}\label{Grigor}

Grigorchuk group $G$ (see e.~g. \cite{Grigorchuk}) can be defined by an automaton with 5 states $a,b,c,d,e$ acting on $X^*=\{0,1\}^*$ as follows:
$$\pi(a,0)=1,\, \pi(a,1)=0, \, \pi(b,x)=\pi(c,x)=\pi(d,x)=x;\, \tau(a,x)=\tau(e,x)=e,$$
$$\tau(b,0)= \tau(c,0)=a,\, \tau(d,0)=e,\, \tau(b,1)=c,\, \tau(c,1)=d,\, \tau(d,1)=b$$
\noindent for any $x\in X$. In particular, $a^2=b^2=c^2=d^2=bcd=e$
(where $e$ denotes the identity). Note that $G$ is not finitely
presented. Observe that the complexity upper bound for the word
problem for $G$ is $O(n\log n)$ \cite{Grigorchuk}. It is known (see
e.~g. \cite{Grigorchuk}) that the portrait property (see
section~\ref{Automaton}) holds for $G$.

%for any word $g\in G$ its portrait of the depth $O(\log|g|)$ defines $g$ defines $g$ uniquely, and all its sections has the length $O(1)$.

In \cite{Lysonok} an algorithm is designed to test whether for given $u,v\in G$ there exists $x\in G$ such that $xux^{-1}=v$. In fact, one can extend this algorithm to produce such $x$, provided it does exist. On the other hand, it seems to be a difficult problem to test whether there exists $x\in G$ such that $xu_ix^{-1}=v_i,\, 1\le i\le m$ for given $u_i,v_i\in G,\, 1\le i\le m$ (and so more, to find such $x$).

One could also use the generalizations $G_{\omega}$
\cite{Grigorchuk84}, \cite{Grigorchuk} of $G$ where $\omega \in
\{0,1,2\}^{\infty}$. Observe that the word problem in $G_{\omega}$
has a complexity upper bound polynomial in the complexity of
computing a prefix of $\omega$ of a logarithmic length, while for a
generic $\omega$ already the single conjugacy equation problem is
more difficult than the similar problem in $G$  \cite{Grigorchuk84},
\cite{Grigorchuk}.

\subsection{Basilica group}\label{basilica}

Consider an automaton group $B$ (sometimes called the Basilica
group) defined by the following automaton with 3 states $a,b,e$
(again, $e$ is the identity of $B$) over the alphabet $X=\{0,1\}$
\cite{Zuk01}, \cite{Zuk02}:
$$\pi(e,x)=\pi(a,x)=x,\, \pi(b,0)=1,\, \pi(b,1)=0;$$
$$\tau(e,x)=\tau(a,0)=\tau(b,0)=e,\, \tau(a,1)=b,\, \tau(b,1)=a$$
\noindent for any $x\in X$.

It is proved in \cite{Zuk01} that the group $B$ also satisfies the
portrait property. Note that for $B$ only an exponential complexity
algorithm is known for the problem of a single conjugacy equation.

%for any word $g\in B$ its portrait of the depth $O(\log|g|)$ defines $g$ defines $g$ uniquely, and all its sections has the length $O(1)$.

\subsection{Universal Grigorchuk group}\label{universal}

One can represent each group $G_{\omega}=F_4/N_{\omega}$ where $N_{\omega}$ is a normal subgroup of 4-free group $F_4$ (with the generators $a,b,c,d$). Denote $N=\bigcap_{\omega} N_{\omega}$ where the intersection ranges over all the infinite words $\omega \in \{0,1,2\}^{\infty}$. The universal group is defined $U=F_4/N$ \cite{Nagnibeda}. Similar to $G$ (see section~\ref{Grigor}) $a^2=b^2=c^2=d^2=bcd=e$ (and again, $U$ is not finitely presented).

One can represent $U$ as an automaton group \cite{Nagnibeda} defined
by an automaton with 5 states   $a,b,c,d,e$ (again, $e$ is the
identity of $U$) over an alphabet $X=\{0,1\}\times \{0,1,2\}$ of
size 6 as follows:
$$\pi(e,(x,y))=\pi(b,(x,y))=\pi(c,(x,y))=\pi(d,(x,y))=(x,y),$$
$$ \pi(a,(0,y))=(1,y),\, \pi(a,(1,y))=(0,y);$$
$$\tau(e,(x,y))=\tau(a,(x,y))=\tau(b,(0,2))=\tau(c,(0,1))=\tau(d,(0,0))=e,$$
$$\tau(b,(0,0))=\tau(b,(0,1))=\tau(c,(0,2))=\tau(d,(0,1))=\tau(d,(0,2))=a,$$
$$\tau(b,(1,y))=b,\, \tau(c,(1,y))=c,\, \tau(d,(1,y))=d$$
\noindent for any $x\in\{0,1\},\, y\in \{0,1,2\}$.

Similar to the group $G$ (cf. section~\ref{Grigor}) the group $U$ also satisfies the portait property \cite{Grigorchuk}, \cite{Nagnibeda}.

%for any word $g\in U$ its portrait of the depth $O(\log|g|)$ defines $g$ defines $g$ uniquely, and all its sections has the length $O(1)$ \cite{Nagnibeda}.

Apparently, the simultaneous conjugacy problem for $U$ (cf.
section~\ref{Anshel}) is not easier than the same problem for $G$,
for $G_{\omega}$ and for $B$.

\subsection{Hanoi 3-Towers group}\label{Hanoi}

We describe Hanoi Towers group $H^{(3)}$ on 3 pegs as an automaton group \cite{Sunik}, \cite{Bondarenko}. The alphabet $X=\{0,1,2\}$ consists of
 3 letters which corresponds to the pegs. Actully, one can generalize to the group $H^{(k)}$ of Hanoi Towers on $k\ge 3$ pegs, then
 $|X|=k$ \cite{Sunik}, \cite{Bondarenko}. A word $x_1\cdots x_n\in X^n$ has a meaning that the disc $i$ is placed on $x_i$-th peg. According to
 the rules of the game in each peg the discs of sizes $1,2,\dots$ are placed in the decreasing order of their sizes from the bottom to the top.

The automaton of $H^{(3)}$ contains 3 states: $a_{01}, a_{02}, a_{12}$. For any word $w\in X^n$ we have
$$a_{ij}(iw)=jw,\, a_{ij}(jw)=iw,\, a_{ij}(xw)=xa_{ij}(w),\, x\not \in \{i,j\}.$$
\noindent This means that $a_{ij}$ takes the disc from the top of either peg $i$ or $j$ being minimal among these two and puts it on another peg among $i$ and $j$. Clearly, $a_{01}^2=a_{02}^2=a_{12}^2=e$ (again, $H^{(k)}$ is not finitely presented).

In \cite{Bondarenko} the portrait property is proved for $H^{(3)}$. Note that the complexity bound
$\exp(O(\log^{k-2}n))$ \cite{Bondarenko} for the word problem in the group $H^{(k)}$ is not polynomial for $k\ge 4$.

\subsection{A group with the unsolvable problem of conjugacy}\label{unsolvable}

In Proposition 7.5 \cite{Bogopolski} a group $F'\subset GL_4(\Z)$ is constructed with generators $M_1,\dots,M_s\in GL_4(\Z)$ having unsolvable orbit problem, i.~e. whether for a pair of vectors $u,v\in \Z^4$ there exists $f\in F'$ such that $fu=v$. In \cite{Ventura} it is proved that the semidirect product $G'=\Z^4 \rtimes F'\subset Aff_4(\Z)$ has the unsolvable conjugacy problem. Moreover, in Proposition 1.5 \cite{Ventura} this construction is modified to make a group $F\subset GL_6(\Z)$ free,  also having the unsolvable orbit problem and $G=\Z^6\rtimes F\subset Aff_6(\Z)$ having the unsolvable conjugacy problem.

On the other hand, the word problem in $G'$ (as well as in $G$) can be solved within the polynomial complexity. Indeed, an element of $G'$ one can represent as a composition of affine transformations in $Aff_4(\Z)$ of $\Z^4$ of the form $v\to u+M_iv,\, 1\le i\le s$ for vectors $u\in \Z^4$. One can explicitly compute such a composition.

Note that in \cite{Ventura} $G$ is represented as an automaton group. It looks reasonable to use both $G$ and $G'$ as  platforms for Anshel-Anshel-Goldfeld scheme (see section~\ref{Anshel}). \vspace{2mm}

{\bf Acknowledgements}. The frst author graciously acknowledges support from
the Simons Foundation through Collaboration Grant 527814, is partially supported by the
mega-grant of the Russian Federation Government (N14.W03.31.0030) and
is grateful to Max-Planck Institut fuer Mathematik, Bonn
during staying in which the paper was conceived. The second author is grateful to the grant RSF 16-11-10075
and to MCCME for inspiring atmosphere.


\begin{thebibliography}{99}

\bibitem{AAG}  I.~Anshel, M.~Anshel, D.~Goldfeld,
An algebraic method for public-key cryptography
, Math. Res.
Lett.
6
(1999) 287--291.

\bibitem{Bartholdi} L.~Bartholdi, R.~Grigorchuk, Z.~Sunik, Branch groups, Handbook of algebra 3, Elsevier (2003) 989--1112.

\bibitem{Nagnibeda} M.~Benli, R.~Grigorchuk, T.Nagnibeda, Universal groups of intermediate growth and their invariant random subgroups,  Funct. Anal. Appl. 49, 3 (2015) 159--174.

\bibitem{Bogopolski} O.~Bogopolski, A.~ Martino, E.~ Ventura,  Orbit decidability and the conjugacy problem for some extensions of groups, Trans. Amer. Math. Soc. 362, 4 (2010),  2003--2036.

\bibitem{Bondarenko} I.~Bondarenko, The word problem in Hanoi Towers groups, Algebra Discr. Math. 17, 2 (2014) 248--255.

\bibitem{Zapata} I.~Bondarenko, N.~Bondarenko, S.~Sidki, F.~Zapata, On the conjugacy problem for finite-state automorphisms of regular rooted trees. With an appendix by R.~Jungers, Groups Geom. Dyn. 7 (2013) 323--355.

\bibitem{Garzon} M.~Garzon, Y.~Zalcstein, The complexity of Grigorchuk group with applications to cryptography, Theor. Comput. Sci. 88 (1991) 83--98.

\bibitem{Grigorchuk84} R.~Grigorchuk, Degrees of growth of finitely generated groups and the theory of invariant means. Math. USSR Izvestiya 25 (1985) 939--985.

\bibitem{Grigorchuk} R.~Grigorchuk, Solved and unsolved problems around one group, Progr. Math. 248, Birkh\'aser (2005) 117--218.

\bibitem{Nekrashevich} R.~Grigorchuk, V.~Nekrashevych, V.~Sushchnskii, Automata, dynamical systems, and groups,  Proc. Steklov Inst. Math. 231 (2000) 128 -203.

\bibitem{Sunik} R.~Grigorchuk, Z.~Sunik, Schreier spectrum of the Hanoi Towers group on three pegs. Analysis on graphs and its applications,  Proc. Sympos. Pure Math. 77, AMS (2008) 183--198.

\bibitem{Wilson} R.~Grigorchuk, J.~Wilson, The conjugacy problem for certain branch groups,
Proc. Steklov Inst. Math. 231 (2000) 204--219.

\bibitem{Zuk01} R.~Grigorchuk, A.~Zuk, Spectral properties of a torsion-free weakly branch group defined by a three state automaton. Computational and statistical group theory,  Contemp. Math. 298, AMS (2002)
57--82.

\bibitem{Zuk02} R.~Grigorchuk, A.~Zuk, On a torsion-free weakly branch group defined by a three state automaton, Int. J. Alg. Comput. 12, 1-2 (2002) 223--246.

\bibitem{GP} D.~Grigoriev, I.~Ponomarenko, Constructions in public-key cryptography over matrix groups, Contemp. Math. 418, AMS  (2006) 103--119.

\bibitem{GS} D.~Grigoriev, V.~Shpilrain, Authentication from matrix conjugation, Groups, Compl., Cryptology  1 (2009) 199--205.

\bibitem{Kahrobaei} M.~Habeeb, D.~Kahrobaei, C.~Koupparis, V.~Shpilrain,
Public key exchange using semidirect
product of (semi)groups
, Lecture Notes Comp. Sci.
7954
(2013), 475--486.

\bibitem{Leonov} Yu.~Leonov, The conjugacy problem in a class of 2-groups, Math. Notes, 64 (1999) 496--505.

\bibitem{Lysonok} I.~Lysonok, A.~Myasnikov, A.~Ushakov, The conjugacy problem in the Grigorchuk group in
polynomial time decidable, Groups, Geom., Dyn. 4 (2010) 813--833.

\bibitem{Menezes} A.~Menezes, P.~van Oorschot, S.~Vanstone,
Handbook of Applied Cryptography,
CRC-Press,
1996.

\bibitem{Myasnikov} A.~Myasnikov, V.~Shpilrain, A.~Ushakov,
Group-based cryptography,
 Birkh\"auser, 2008.

\bibitem{Ushakov} A.~Myasnikov, A.~Ushakov, Cryptanalysis of the Anshel-Anshel-Goldfeld-Lemieux Key Agreement Protocol, Groups, Compl., Cryptology 1, 1 (2009) 63--76.

\bibitem{Petrides} G.~Petrides, Cryptanalysis of the public key cryptosystem based on the word problem on the Grigorchuk groups. Lect. Notes Comput. Sci. 2898, Springer (2003) 234--244.

\bibitem{Rozhkov} A.~Rozhkov, The conjugacy problem in an automorphism group of an infinite tree, Math. Notes 64 (1999) 513--517.

\bibitem{Ventura} Z.~Sunik, E.~Ventura, The conjugacy problem in automaton groups is not solvable,  J. Algebra 364 (2012) 148--154.
        %
        %\bibitem{AGG} M. Akian, S. Gaubert, A. Guterman, Tropical polyhedra are equivalent to mean payoff games,
        %Internat. J. Algebra Comput., 22, 1 (2012), 43 pp.
        %
        %\bibitem{BJS} T. Bogart, A. N. Jensen, D. Speyer, B. Sturmfels, R.R. Thomas,
        %Computing tropical varieties, J. Symbolic Comput. 42, 1--2 (2007) 54--73.
        %
        %\bibitem{BH} P. Butkovic, G.Heged\"us, An elimination method for finding all solutions of the system
        %of linear equations over an extremal algebra, Ekon.-Mat. Obzor 20 (1984) 203--214.
        %
        %\bibitem{C86} A. Chistov, An algorithm of polynomial complexity for factoring polynomials,
        %and determination of the components of a variety in a subexponential
        %time, J.Soviet Math., 34 (1986) 1838--1882.
        %
        %\bibitem{C86-2} A. Chistov, Polynomial complexity of Newton-Puiseux algorithm, Lect.
        %Notes Comput. Sci., 233 (1986) 247--255.
        %
        %\bibitem{DSS} M. Develin, F. Santos, B. Sturmfels, On the rank of a tropical matrix, In:
        %Combinatorial and Computational Geometry, MSRI Publications, v. 52 (2005).
        %
        %\bibitem{DS} M. Develin, B. Sturmfels, Tropical convexity, Doc. Math., 9 (2004) 1--27.
        %
        %\bibitem{DW} A. Dress, W. Wenzel, Algebraic, tropical, and fuzzy geometry, Beitr. Algebra Geom. 52, 2 (2011) 431   -461.
        %
        %\bibitem{GK} S. Gaubert, R.D. Katz, Minimal half-spaces and external representation of tropical polyhedra,
        %J. Algebraic Combin. 33, 3 (2011) 325   -348.
        %
        %\bibitem{G86} D. Grigoriev, Polynomial factoring over a finite field and solving systems
        %of algebraic equations, J. Soviet Math., 34 (1986) 1762--1803.
        %
        %\bibitem{G13} D. Grigoriev, Complexity of solving tropical linear systems, Computational Complexity, 22 (2013) 71--88.
        %
        %\bibitem{G15} D. Grigoriev, Polynomial complexity recognizing a tropical linear variety,  Lect.
        %Notes Comput. Sci., v. 9301 (2015) 152--157.
        %
        %\bibitem{GP} D. Grigoriev, V. Podolskii,
        %Complexity of tropical and min-plus linear prevarieties, Computational Complexity, 24, 1 (2015) 31--64.
        %
        %\bibitem{GV} D. Grigoriev, N. Vorobjov, Upper bounds on Betti numbers of tropical prevarieties,
        %Arnold Math. J., online first (2018).
        %
        %\bibitem{HT} K. Hept, T. Theobald, Tropical bases by regular projections, Proc. Amer. Math. Soc., 137, 7 (2009)
        %2233--2241.
        %
        %\bibitem{JMM} A. Jensen, H. Markwig, T. Markwig, An algorithm for lifting points in a tropical variety,
        %Collect. Math., 59, 2 (2008) 129--165.
        %
        %\bibitem{JY} A. Jensen, J. Yu, Stable intersections of tropical varieties, Journal of Algebraic Combinatorics,
        %43, 1 (2016) 101--128.
        %
        %\bibitem{J} M. Joswig, The Cayley trick for tropical hypersurfaces with a view toward Ricardian economics,
        %in: Homological and Computational Methods in Commutative Algebra, Springer INdAM Series, 20, Springer, 2017.
        %
        %\bibitem{KK}  Ya. Kazarnovskii, A. G. Khovanskii, Tropical noetherity and Gr\"obner bases,
        %St. Petersburg Math. J., 26, 5 (2015) 797--811.

%        \bibitem{B}  L. E. J. Brouwer, Zur Invarianz des n-dimensionalen Gebiets, Mathematische Annalen 72, (1912) 55--56.

 %       \bibitem{C} F.~H.~Clarke, On the inverse function theorem, Pacific Journal of Mathematics, 64, 1 (1976) 97--102.

  %      \bibitem{G} D.~Grigoriev, Tropical Newton-Puiseux polynomials, Lect. Notes Comput. Sci., 11077 (2018) 177--186.

   %     \bibitem{HK} S.~Hencl, P.~Koskela, Lectures on mappings of finite distortion,
    %    Lecture Notes in Mathematics, 2096. Springer, Cham, 2014.

     %   \bibitem{H} C.-W.~Ho, A note on Proper Maps, Proc. AMS, 51, 1 (1975) 237--241.

      %  \bibitem{MS} D.~Maclagan, B.~Sturmfels, Introduction to Tropical Geometry, American Math. Society, 2015.

       % \bibitem{OR} E.~Outerelo, J.~M.~Ruiz, Mapping Degree Theory, American Math. Society, 2009.

        %\bibitem{R} D.~Radchenko, Approximation by maps with nonnegative Jacobian, Math. Notes, 93, 1-2 (2013) 297--307.

        %\bibitem{M-I} G.~G.~Magaril-Il'aev, The implicit function theorem for Lipschitz maps, Russian Math. Surveys 33:1 (1978) 209-210.

        %\bibitem{OP} B. Osserman, S. Payne, Lifting tropical intersections, Documenta Mathematica, 18 (2013) 121--175.
        %
        %\bibitem{RST} J. Richter-Gebert, B. Sturmfels, T. Theobald, First steps in tropical geometry, In:
        %G. Litvinov, V. Maslov (Eds.), Idempotent Mathematics and Mathematical Physics
        %(Proceedings Vienna 2003), Contemporary Mathematics, 377, American Math. Society (2005) 289--317.
        %
        %\bibitem{S} D. Speyer, Tropical linear spaces, SIAM J. Discrete Math., 22, 4 (2008) 1527   -1558.
        %
        %\bibitem{YY} J. Yu, D.S. Yuster, Representing tropical linear spaces by circuits,
        %The 19th International Conference on Formal Power Series and Algebraic Combinatorics, 2007,
        %arXiv:0611579, 2006.

        %\bibitem{ST} R. Steffens, T. Theobald, Combinatorics and genus of tropical intersections and Ehrhart theory, %SIAM J. Discrete Math. 24 (2010) 17--32.

    \end{thebibliography}
\end{document}